\documentclass[12pt]{cicp}
\usepackage[utf8]{inputenc}
\usepackage[english]{babel}
\usepackage[dvips]{epsfig}
\usepackage{graphics,graphicx}
\begin{document}

\title[Contact angles in multicomponent LB models]{Contact angle
determination in multicomponent lattice Boltzmann simulations}

\author[S. Schmieschek, J. Harting]{Sebastian Schmieschek\affil{1,2} and Jens Harting\affil{2,1}\corrauth}

\address{
\affilnum{1}\ Institute for Computational Physics, University of Stuttgart,\\
Pfaffenwaldring 27, 70569 Stuttgart, Germany\\
\affilnum{2}\ Department of Applied Physics, Eindhoven University of Technology,\\ 
P.O. Box 513, 5600 MB Eindhoven, The Netherlands.
}

\emails{{\tt sschmie@icp.uni-stuttgart.de} (S.~Schmieschek), {\tt j.harting@tue.nl} (J.~Harting)}

\begin{abstract}
Droplets on hydrophobic surfaces are ubiquitous in microfluidic applications and
there exists a number of commonly used multicomponent and multiphase lattice
Boltzmann schemes to study such systems. In this paper we focus on a popular
implementation of a multicomponent model as introduced by Shan and Chen. Here,
interactions between different components are implemented as repulsive forces
whose strength is determined by model parameters. In this paper we present
simulations of a droplet on a hydrophobic surface. We investigate the
dependence of the contact angle on the simulation parameters and quantitatively
compare different approaches to determine it. Results show that the method is
capable of modelling the whole range of contact angles. We find that the
a priori determination of the contact angle is depending on the
simulation parameters with an uncertainty of 10 to 20\%.  
\end{abstract}

\pac{
47.55.D-, 
47.11.-j 
}


\maketitle

\section{Introduction}
During the last few decades the miniaturization of technical devices down
to submicrometric sizes has made considerable progress. In particular,
during the 1980s, so-called microelectro-mechanical systems (MEMS) became
available for chemical, biological and technical applications leading to
the rise of the discipline called ``microfluidics'' in the
1990s~\cite{bib:tabeling-book}. In microfluidic devices the surface to
volume ratio of a fluid can be large and thus a good understanding of the
behavior of the fluid close to the surface is mandatory. However, the
behavior of a fluid close to a solid interface is very complex and
involves the interplay of many physical and chemical properties. These
include the wettability of the solid, the shear rate or flow velocity, the
bulk pressure, the surface charge, the surface roughness, as well as
impurities and dissolved gas.

A common concept to quantify the wettability of a surface is the so called
contact angle. The contact angle is the angle at which the interface
between a liquid and a gas or vapor meets a solid surface. If the contact
angle is larger than 90$^{\circ}$, the surface is called non-wettable
(hydrophobic if the liquid is water) and if the angle is smaller than
90$^{\circ}$, it is said to be wettable (hydrophilic).
Superhydrophobic surfaces are surfaces with contact angles larger than
150$^{\circ}$. Here, almost no contact between droplet and surfaces can be
observed and the effect is often referred to as ``Lotus effect''.
Regardless of the amount of wetting, the shape of the drop can be
approximated by a truncated sphere.

\begin{figure}[h]
\centering
\includegraphics[width=0.6\textwidth]{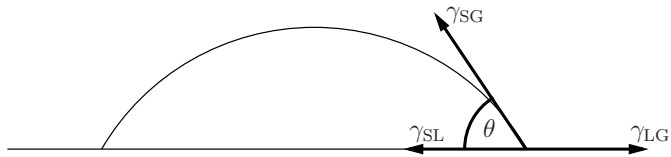}
\caption{Definition of the contact angle as given by Young's
equation.\label{fig:droplets}}
\end{figure}

For a droplet on an idealised smooth surface, the contact angle $\theta$ can be
computed using the surface tensions between liquid and gas
$\gamma_{\rm LG}$, liquid and surface $\gamma_{\rm LS}$ and surface and gas
$\gamma_{\rm SG}$ as given
by Young's equation~\cite{Young1805} (see Fig.~\ref{fig:droplets}),
\begin{equation}
\cos \theta = \frac{\gamma_{\rm SG} - \gamma_{\rm SL}}{\gamma_{\rm LG}} .
\label{eq:young}
\end{equation}

The model of Young was extended by Wenzel~\cite{Wenzel1949} as well as
Cassie and Baxter~\cite{Cassie44} in order to take the influence of
surface roughness into account. While Wenzel describes a state where the
surface is completely covered by the liquid, Cassie and Baxter describe a
state where gas bubbles are enclosed between the liquid and the rough
surface. Both states have been observed experimentally and in
simulations~\cite{bib:jens-jari:2008,Yeh08}.
The transition between the Wenzel and the Cassie-Baxter state leads to the
phenomenon of contact angle hysteresis as observed for droplets on a
tilted surface where one has to distinguish between the advancing and the
receding contact angle~\cite{Dorrer07,Pirat08,Kusumaatmaja08}.
In particular the state proposed by Cassie and Baxter is of technological
interest since it can be used to significantly increase the contact angle
in order to generate superhydrophobic
surfaces with $\theta > 150^{\circ}$~\cite{Vandembroucq97,Marmur08,Roach08}.
Such surfaces can be utilized to increase the flow velocity and thus the
mass flux in microchannels~\cite{Bouzigues08,bib:jens-kunert:2008c}.

While both molecular dynamics and lattice Boltzmann methods (LBM) have been employed to simulate systems with wetting properties, only LBM allow to reach experimentally relevant time- and length scales. Therefore,
the method has become very popular to simulate typical problems occurring
in microfluidics. A particular advantage of the lattice Boltzmann approach
is the availability of established multiphase or multicomponent
methods~\cite{bib:shan-chen-93,bib:shan-chen-liq-gas,bib:swift-osborn-yeomans,bib:swift-orlandini-osborn-yeomans,bib:gunstensen-rothman-zaleski-zanetti}
and a straight forward implementation of complex boundary conditions. This
allows the simulation of multiphase or multicomponent fluid flow along
interacting surfaces~\cite{bib:jens-kunert-herrmann:2005,bib:jens-kunert:2007b,bib:jens-kunert:2008b,bib:jens-kunert:2008c}.
While the free energy based multiphase model introduced by Swift et
al.~\cite{bib:swift-osborn-yeomans} allows to set the contact angle
directly, this possibility does not exist for the model introduced by Shan
and Chen. Here, the surface tension and thus the contact angle only
appear indirectly by tuning the interaction between different fluid
species and the surface~\cite{bib:shan-chen-93,bib:shan-chen-liq-gas}.
Therefore, a proper determination of the contact angle is of fundamental
importance for reliable comparisons between simulation results and those
obtained from theory and experiment. For the single component multiphase
Shan-Chen model, Benzi et al. proposed an analytical ansatz to compute the
contact angle~\cite{benzi-etal-06b}. However, in this paper we focus on
the multicomponent model~\cite{bib:shan-chen-93} and restrict ourselves to
single phase components only. For such a model, Huang et
al.~\cite{bib:huang-thorne-schaap-sukop-2007} recently proposed an
estimate determining the contact angle. However, a full analytical
solution of the problem is still missing. Therefore, we compare and
discuss different methods to quantify $\theta$ in dependence on the
parameters of the simulation model, namely the geometrical measurement,
the approach of Huang et al., as well as utilizing measurements of the
surface tension to solve Young's equation.

\section{Simulation method}
A set of equations can be used to represent a standard lattice Boltzmann system involving
multiple species~\cite{bib:higuera-succi-benzi}
\begin{equation}
\label{LBeqs}
n_k^{\alpha}({\bf x}+{\bf c}_k, t+1) - n_k^{\alpha}({\bf x},t) = 
\Omega_k^{\alpha},
\end{equation}
with $k= 0,1,\dots,b$.
The single-particle velocity distribution function $n_k^{\alpha}({\bf x},t)$
indicates the density of species $\alpha$, having velocity ${\bf c}_k$, at
site ${\bf x}$ on a
D-dimensional lattice of coordination number $b$, at timestep $t$. The
collision operator
\begin{equation}
\Omega_k^{\alpha}=-\frac{1}{\tau^{\alpha}}(n_k^{\alpha}({\bf x},t) -
n_k^{\alpha,r eq}({\bf x},t)),
\end{equation}
represents the change in the
single-particle distribution function due to the collisions. A popular
form is the single relaxation time $\tau_{\alpha}$, linear `BGK'
form~\cite{bib:bgk} for the collision operator.  It can be shown
for low Mach numbers that the LB equations correspond to a solution of the
Navier-Stokes equation for isothermal, quasi-incompressible fluid flow.
The lattice Boltzmann method is an excellent candidate to exploit the
possibilities of parallel computers, as the computations a lattice site
require only information about quantities at nearest neighbour lattice
sites~\cite{bib:love-nekovee-coveney-chin-gonzalez-martin,bib:jens-harvey-chin-venturoli-coveney:2005}.
The
local equilibrium distribution $n_k^{\alpha,eq}$ plays a fundamental role
in the dynamics of the system as shown by Eq.~(\ref{LBeqs}). In this
study, we use a purely kinetic approach, for which
$n_k^{\alpha,eq}({\bf x},t)$ is derived by imposing certain
restrictions on the microscopic processes, such as explicit mass and global
momentum conservation~\cite{bib:chen-chen-martinez-matthaeus}
\begin{equation}
\label{Equil}
n_k^{\alpha,eq} = 
 \zeta_k\rho^{\alpha}\left[1+
\frac{{\bf c}_k {\bf u}}{c_s^2} +\frac{({\bf c}_k {\bf u})^2}{2c_s^4}
-\frac{u^2}{2c_s^2}+\frac{({\bf c}_k {\bf u})^3}{6c_s^6}
-\frac{u^2({\bf c}_k {\bf u})}{2c_s^4}\right],
\end{equation}
where
$\rho^{\alpha}({\bf x},t)\equiv\sum_k \eta_k^{\alpha}({\bf x},t)$ is the
fluid density and
${\bf u} = {\bf u}({\bf x},t)$ is the macroscopic bulk velocity
of the fluid, given by $\rho^{\alpha}({\bf x},t){\bf u}^{\alpha}
\equiv \sum_k n_k^{\alpha}({\bf x},t){\bf c}_k$. $\zeta_k$ are the
coefficients resulting from the velocity space discretization and
$c_s$ is the speed of sound, both of which are determined by the
choice of the lattice. We use a D3Q19 implementation, i.e., a three
dimensional lattice with 19 discrete velocities.
Immiscibility of species $\alpha$ is introduced in the model following
Shan and Chen~\cite{bib:shan-chen-93,bib:shan-chen-liq-gas}, where only
nearest neighbour
interactions among the species are considered.  These
interactions are described by a self-consistently generated mean field
body force
\begin{equation}
\label{Eq:SCforce}
{\bf F}^{\alpha}({\bf x},t) \equiv -\psi^{\alpha}({\bf x},t)\sum_{\bf
\bar{\alpha}}g_{\alpha \bar{\alpha}}\sum_{\bf
x^{\prime}}\psi^{\bar{\alpha}}({\bf x^{\prime}},t)({\bf x^{\prime}}-{\bf
x})\mbox{ ,}
\end{equation}
where $\psi^{\alpha}({\bf x},t)$ is the so-called effective mass, which
can have a general form for modeling various types of fluids (we use
$\psi^{\alpha} = (1 - e^{-\rho^{\alpha}}$)\cite{bib:shan-chen-93}), and
$g_{\alpha\bar{\alpha}}$ is a force coupling constant whose magnitude
controls the strength of the interaction between components $\alpha$,
$\bar{\alpha}$ and is set positive to mimic repulsion. The symbol ${\bf x^{\prime}} = 
{\bf x} + {\bf c}_{k}$ denotes the position of a nearest neighbour. The dynamical
effect of the force is realized in the BGK collision operator by adding the increment
\begin{equation}
\delta{\bf u}^{\alpha} = \frac{\tau^{\alpha}{\bf
F}^{\alpha}}{\rho^{\alpha}}
\end{equation}
 to the velocity ${\bf u}$ in the equilibrium distribution (Eq.~(\ref{Equil})). 
This naturally opens the way to introduce similar
interactions between each fluid species and the channel walls, where the
strength of the interaction is determined by the fluid densities, free
coupling constants, and a wall interaction parameter.

For the interaction of the fluid components with the channel walls we
apply mid-grid bounce back boundary conditions~\cite{bib:succi-01} and assign
interaction properties to the wall which are similar to those of an
additional fluid species, i.e. we specify constant values for the force
coupling constant $g_{\bar{\alpha}\alpha}=g_{\rm wall,\alpha}$ and the density
$\eta^{\bar{\alpha}}=\eta_{\rm wall}$ for the rest vector ($c_k=0$, $k=0$) at wall boundary nodes of the lattice.
This results in a purely local force as given in Eq.~\ref{Eq:SCforce} between
the flow and the boundaries. Even though one could argue that a single
parameter to tune the fluid-wall interaction would be sufficient, we keep our
approach as close as possible to the original idea of Shan and Chen in order to
benefit from the experience obtained from other works using the original model.
Furthermore, the additional parameter allows more flexibility to tune the
interactions in a system more complex than considered here.  The fluid-wall
interaction can be linked to a contact angle between fluid droplets and solid
walls as it is often used to quantitatively describe hydrophobic
interactions~\cite{bib:deGennes-85}. 

In the model used, the interface between domains of different fluid
species has a finite width. In order to define a position of an interface 
we introduce the order parameter $\phi =
\rho^{\alpha}-\rho^{\bar{\alpha}}$ which is zero at the interface.

We perform simulations of a droplet at an interacting surface in order to
investigate the influence of the droplet size, the pseudo wall density
(wettability) $\eta_{\rm wall}$, and the coupling constant
$g_{\alpha\bar{\alpha}}$ on the resulting contact angle.
The system is initialised with a spherical cap of component $A$ and
density $\rho^{A} = 0.7$ at a smooth surface. The drop is surrounded by a
fluid of component $B$ and density $\rho^{B}=0.7$. 

This choice of densities is made without loss of generality. In the scope
of this work only one coupling parameter $g_{\bar{\alpha}\alpha}$ is used. Introduction
of a density contrast at initialisation therefore mainly results in a shift in 
droplet size and mean density. The equilibrium density contrast, however, is 
fixed by the Laplace law.
To quantitatively describe a droplet of fluid in a gaseous medium, typically a 
contrast in dynamic viscosities of the order of $10^{3}$ needs to be modelled. 
This is well beyond the limit of numerical stability of the model employed.
Despite this fact, as shown below, the phenomenological nature of the Shan-Chen force allows the
qualitative modelling of the whole contact angle range.

At the surface mid-grid bounce back boundary conditions as well as a repulsive
force with pseudo wall density $\eta_{\rm wall}$ are applied.

\section{Geometrical determination of the contact angle}

\begin{figure}[ht]
\centering
\includegraphics[width=0.45\textwidth]{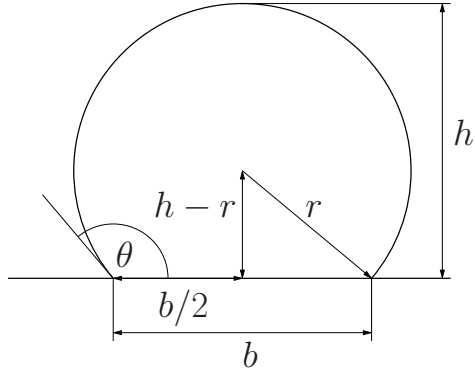}
\caption{Geometrical measurement of the contact angle. The contact angle
can be determined by measuring the diameter of the base $b$ and the height
$h$. The radius of the droplet is given by $r=\frac{4h^{2}+b^{2}}{8h}$.\label{fig:geoCA1}}
\end{figure}

Assuming a droplet has the shape of a spherical segment, the contact angle
\begin{equation}\label{eq:geoCA}
\theta = \pi - \arctan \frac{b/2}{r-h}
\end{equation}
can be obtained by measuring the base $b$, the height $h$ and the radius
$r$ of the droplet (see Fig.~\ref{fig:geoCA1}). The geometrical measurement is
used as a reference to compare to the approaches of contact angle determination
further below. Base and height can be determined by measuring the position where 
the order parameter has a value of zero. The radius is then given by 
$r=\frac{4h^{2}+b^{2}}{8h}$. 
Due to the fluid-wall interaction there exists an interface layer in the vicinity
of the wall. Determining the base by measurement of sign change of the order 
parameter immediately above the wall is therefore introducing an error. To
avoid this, the droplet radius is calculated from the base and height relative
to a reference point sufficiently far from the interface layer. For the
simulation results discussed here a height-offset of 5 lattice units proved to be
sufficient. The correct base length above the wall can then be calculated from 
the so-determined radius and the actual height.

\section{Dependence of the contact angle on model parameters}
\label{subsec:parms}
\begin{figure}[ht]
\includegraphics[width=0.585\textwidth]{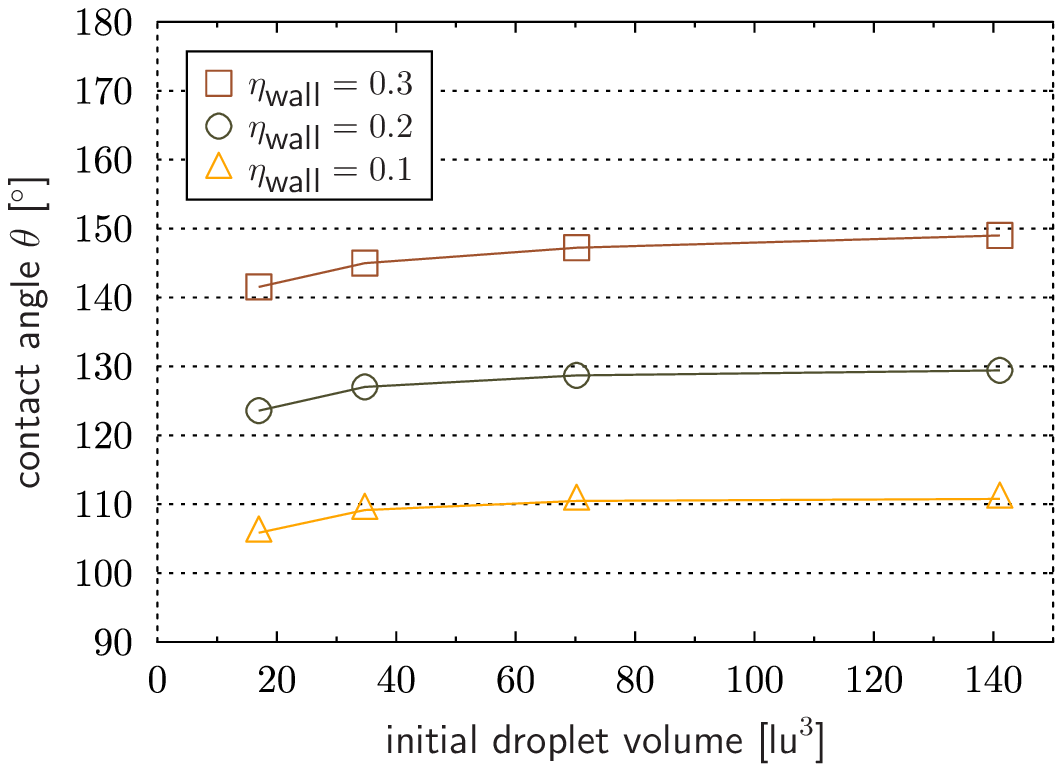}
\hspace{-0.1\textwidth}
\includegraphics[width=0.585\textwidth]{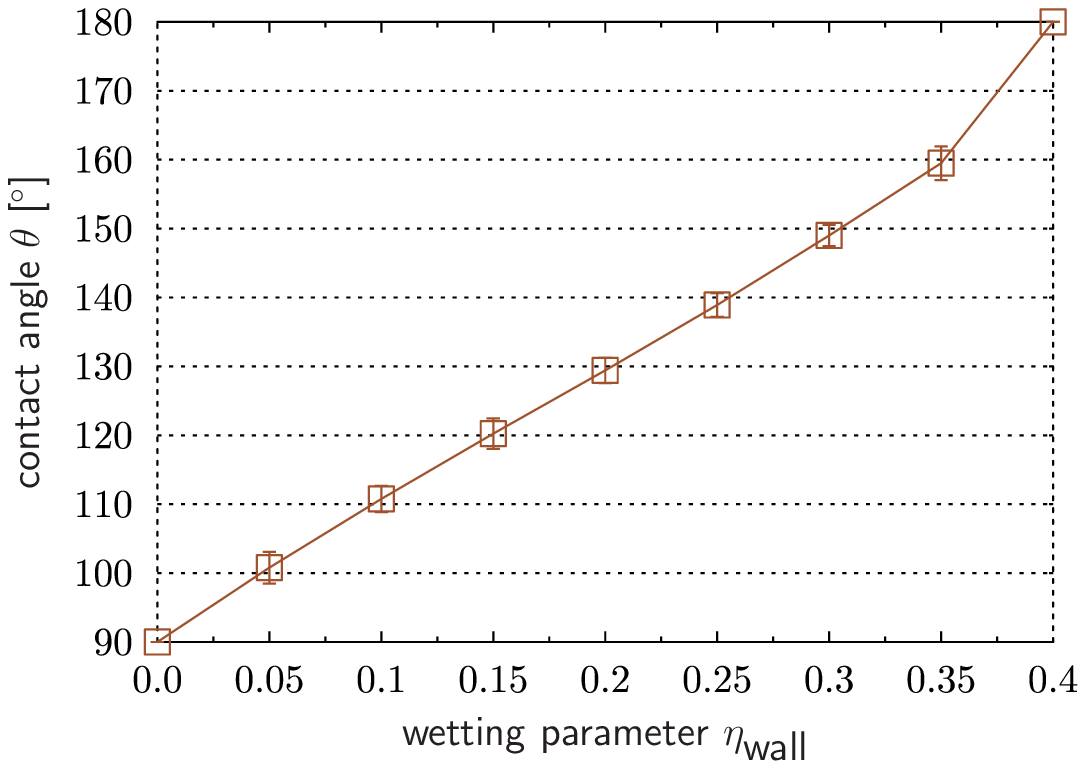}
\caption{Left: dependence of $\theta$ on the droplet size for different
fluid-surface interactions. $g_{\alpha\bar{\alpha}}$ is kept fixed at
0.16.
Right: contact angle versus wetting parameter $\eta_{\rm wall}$ for a
droplet with initial volume $141^3$ and $g_{\alpha\bar{\alpha}}=0.16$. The
error bars denote values obtained from assuming the interface position being given by
half the maximum absolute value of the order parameter $\phi$.
Lines drawn are a guide to the eye.\label{fig:geoCA2}}
\end{figure}

The size of the simulated system has a strong influence on the precision
of the results. For example, due to discretization effects, a droplet
cannot be approximated by a sphere if the lattice resolution is too low.
Further, calculations that take the curvature of the drop into account, 
also require a well resolved surface of the droplet. 
\begin{figure}[h]
\centering
\includegraphics[width=0.7\textwidth]{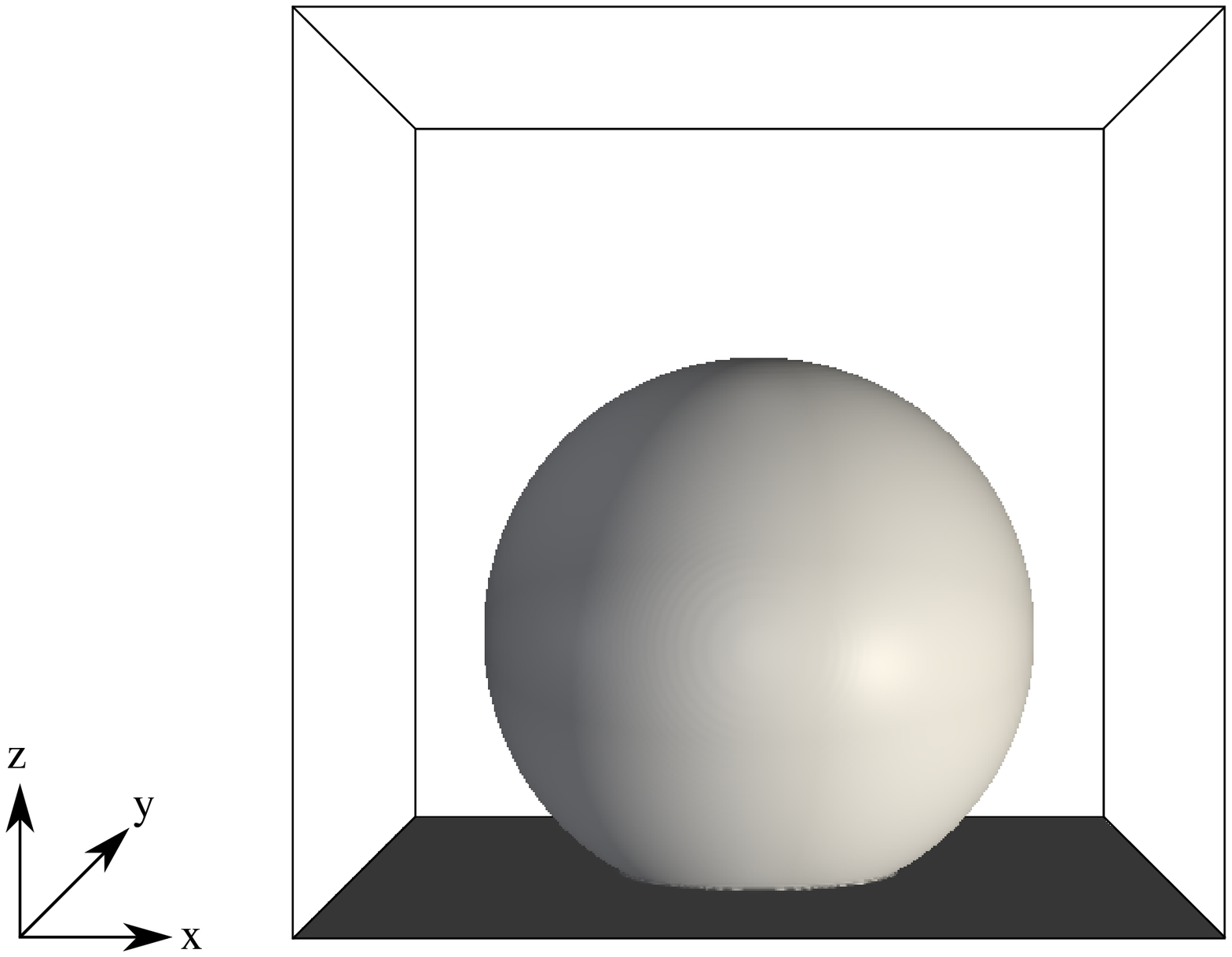}
\caption{Visualisation of simulation data. The black box indicates the
simulated volume. Only the liquid component of the droplet and the wall
component are rendered. Apart from the solid boundary at the bottom periodic
boundaries were employed. Here the system size is $256^3$ lattice sites,
corresponding to an initial droplet volume of  $141^3$ lattice sites. The
wetting parameters are $\eta_{\rm wall}=0.3$ and $g_{\alpha\bar{\alpha}}=0.16$.
}\label{fig:3Ddrop}. 
\end{figure}
We checked the dependence of the contact angle on the system size for $32^{3}$,
$64^{3}$, $128^{3}$, and $256^{3}$ lattices and initial droplet volumes of
$17^{3}$, $35^{3}$, $70^{3}$, and $141^{3}$. An example system setup is shown
in Fig.~\ref{fig:3Ddrop}. The left side of Fig.~\ref{fig:geoCA2} depicts the
measured contact angle in a system of two immiscible components of equal
density $\rho=0.7$ and kinematic viscosity $\nu=(2\tau-1)/6=1/6$ for
$g_{\alpha\bar{\alpha}}=0.16$ and $\eta_{\rm wall}=0.1, 0.2, 0.3$. It can be
seen that the contact angle increases with increasing absolute value of
$\eta_{\rm wall}$ and that even for the largest system size the contact angle
is not fully converged. The convergence depends on the wettability parameter
$\eta_{\rm wall}$: the stronger the repulsion between fluid and surface, the
larger the system has to be. However, considering the doubling of initial
droplet volume from $\approx70^3$ $l.u.^{3}$ to $\approx141^3$ $l.u.^{3}$ the
relative change in the contact angle measured is as low as approximately
$0.21\%$ for $\eta_{\rm wall}=0.1$, $0.56\%$ for $\eta_{\rm wall}=0.2$ and
$1.2\%$ for $\eta_{\rm wall}=0.3$.  Therefore, we find a compromise between
optimal use of computing time and precision of the measurement and restrict
ourselves to lattices of size $128^{3}$, and $256^{3}$.  Figure
\ref{fig:geoCA2} (right) shows the dependence of the contact angle on the
wetting parameter $\eta_{\rm wall}$ for $g_{\alpha\bar{\alpha}}=0.16$. The plot
shows a linear dependence of the contact angle up to about $160^{\circ}$ and
$\eta_{ {\rm wall}}=0.35$. In the vicinity of the complete dewetting limit,
the dependence becomes non-linear.

\subsection{Surface tension measurements at planar interfaces}
As given by Eq.~\ref{eq:young}, the contact angle can be calculated if the
surface tensions between liquid and gas, liquid and surface, and gas and
surface are known. Only the curvature of the interface between liquid and
gas depends on the size of the droplet. By assuming
an infinitely large droplet on a surface,
the interface between liquid and gas can be approximated as planar and
the surface tension can be calculated using its mechanical definition
\begin{equation}
\label{eq:gammaMechDef}
\gamma = \int_{-\infty}^{\infty} P_{N}-P_{T} \quad {\rm d}x,
\end{equation}
wherein the component of the pressure tensor normal to the interface is $P_{N}=P_{zz}$ 
and the component transversal to the interface is $P_T=P_{xx}=P_{yy}$. 
The pressure tensor is computed as 
\begin{eqnarray}
  P_{ij}(\mathbf{x},t)
  &\equiv&
  \sum_{\alpha}
  \sum_k
  \Big(\mathbf{c}_{ki}-\mathbf{u}_{i}(\mathbf{x},t)\Big)
  \Big(\mathbf{c}_{kj}-\mathbf{u}_{j}(\mathbf{x},t)\Big)
  n_k^\alpha(\mathbf{x},t)		
  \nonumber\\
  &+&
  \frac{1}{4} \sum_{\alpha,\bar{\alpha}}
  g_{\alpha\bar{\alpha}}  
  \sum_{\mathbf{x}^\prime} 
  \Big[\psi^\alpha(\mathbf{x})
  \psi^{\bar{\alpha}}(\mathbf{x}^\prime) 
  + 
  \psi^{\bar{\alpha}}(\mathbf{x},t)
  \psi^{\alpha}(\mathbf{x}^\prime,t)\Big] 
  (\mathbf{x}-\mathbf{x}^\prime)^{2}.
\end{eqnarray}
Here, the first term is equivalent to the dynamic pressure. The second term describes 
the distribution of the mean field body force given by eq.~\ref{Eq:SCforce}.

For interfaces between liquid or gas and the surface, $\gamma$ is being
computed equivalently. As introduced in
Ref.~\cite{bib:gonzalez-coveney-04} a $8\times 8\times 128$ sized system with periodic
boundaries is filled with two 64 lattice units long lamellae of different
fluid components. The densities for both components are chosen as $0.7$,
$g_{\alpha\bar{\alpha}}$ is varied between $0.0$ and $0.2$ in steps of
0.02. For calculating the surface tension between a fluid component and
the wall, half of the system is filled with a wall component with variable
wetting parameter $\eta_{\rm wall}$ between $0.0$ and $0.6$ in steps
of $0.02$. $g_{\alpha\bar{\alpha}}$ is varied as for the fluid-fluid
case.

\begin{figure}[h]
\centering
\includegraphics[width=0.7\textwidth]{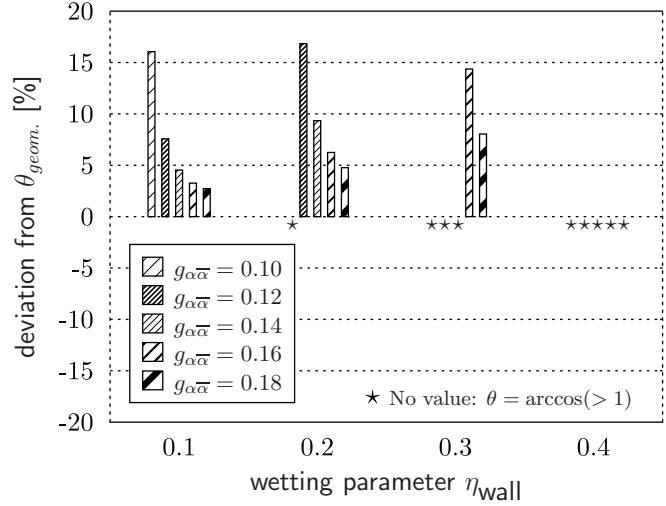}
\caption{Deviation of contact angles obtained from measurements at planar
interfaces from values obtained from geometrical measurements at a droplet
on a surface.
}\label{fig:plaCAreldiff}. 
\end{figure}

The surface tension obtained is being used to calculate the contact angle
as given by Eq.~\ref{eq:young}. Figure~\ref{fig:plaCAreldiff} shows the 
deviation of the obtained values from the ones obtained by a geometrical
determination of $\theta$. 
For the geometrical measurements, a droplet with a volume of $70^3$ $l.u.^3$
on a flat surface is used.
It can be seen that the deviation is always
positive and that the dependence of $\theta$ on the model parameters is
stronger than for the geometric measurements. In fact, already
$g_{\alpha\bar{\alpha}} = 0.10$ and values for $\eta_{\rm wall}$ of
$0.2$ cause the contact angle to reach $180^{\circ}$, while for
$g_{\alpha\bar{\alpha}} = 0.18$ this value is not being reached for
$\eta_{\rm wall}=0.3$. For $\eta_{\rm wall}=0.4$ all simulations
have produced contact angles of $180^{\circ}$. 
 
The significant differences between the geometrical determination of the
contact angle and the measurements of the surface tension have a number of
reasons: 
first, fluids diffuse into areas where the other component is the
majority. Thus, in the droplet system, the volume covered by the droplet
also includes up to 5\% of the surrounding fluid component which has an
influence on the measured surface tension. Further, since the pressure is
tensorial at the interface only, merely seven discrete data points along one
axis are used to calculate the surface tension. Enhancing the resolution of the
interface would, however, increase the computational cost significantly.
Nonetheless, the measurement might be improved by introducing better statistics
interpolating over the whole droplet interface.
The fluid components are slightly compressible leading to slightly different
maximum and minimum values of the steady state densities of the droplet
system and the planar setup for the surface tension determination.
Further, the curvature of the interface is not being taken into account.
In particular for small droplets, this effect has a significant
influence. Therefore, we compare our results to measurements obtained
using an equation for the surface tension that takes the droplet geometry
into account:
\begin{equation}\label{eqn:mechRadGamma} \gamma = \int_{0}^{\infty} \left(
\frac{r}{R_{s}} \right)^{2} (P_{N}-P_{T}) {\rm d}r
\end{equation}
Here $R_{s}$ is the radius of the interface. We integrate from the center of
the droplet ($r=0$). The integral is evaluated until 5 l.u. before the
border of the system in order to minimize any influences due to periodic
boundary conditions.

\begin{figure}[h]
\centering
\includegraphics[width=0.7\textwidth]{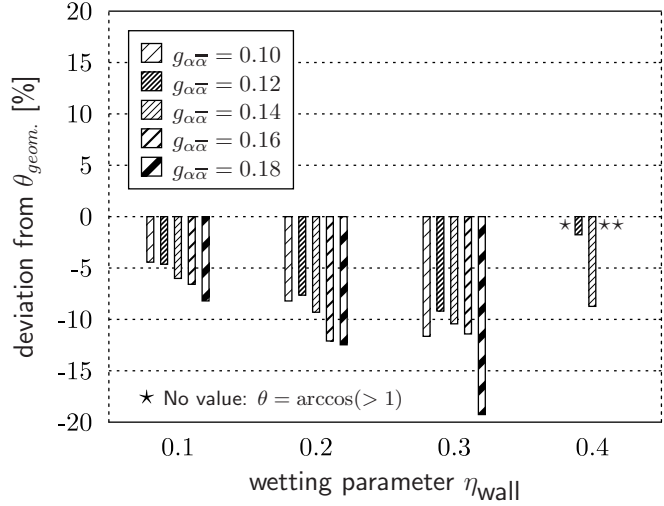}
\caption{Deviation (in percent) of the contact angle measured using
Young's equation and surface tensions obtained from radial interfaces of a
droplet system from geometrical measurements. The initial volume of the
droplet is $70^3$ $l.u.^3$.\label{fig:radCAreldiff}}
\end{figure}

The resulting contact angles are always smaller than the ones obtained
from the geometrical determination (see Fig.~\ref{fig:radCAreldiff}).  In
particular for moderate values of $\eta_{\rm wall}$ we find strong
deviations due to a higher curvature in the curve solving the Young-Laplace equation with the measured surface tensions.
There is no linear dependence of the contact angle on the surface wettability as observed in the geometrical measurements.

\begin{figure}[h]
\centering
\includegraphics[width=0.7\textwidth]{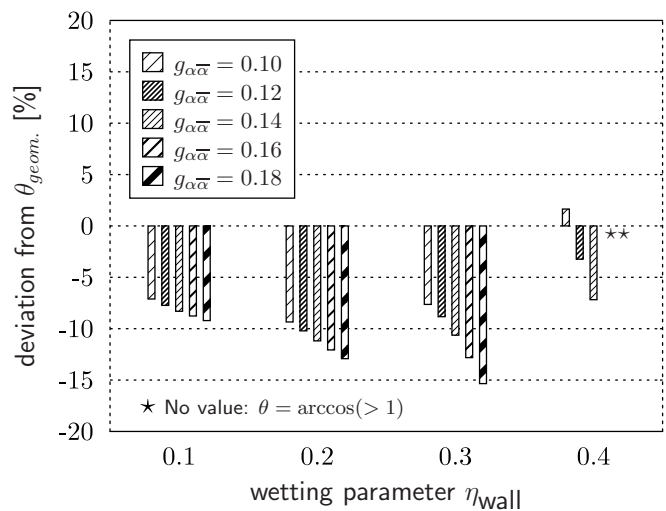}
\caption{Deviation (in percent) of the contact angle measured using the
approach of Huang at al. from geometrical measurements. The initial volume
of the droplet is $70^3$ $l.u.^3$.\label{fig:hngCAreldiff}}
\end{figure}

In a recent publication, Huang et al. postulate an estimate for the
contact angle within the multiphase multicomponent Shan-Chen
model~\cite{bib:huang-thorne-schaap-sukop-2007}. This estimate is valid for a fixed ratio
of the component densities and their coupling constants.
The approach of Huang et al. is based on the assumption that the surface
tension at the wall is mainly determined by the local interaction.
The force acting on component $\alpha$, where the boundary condition is
given by an interacting surface, can be written as
$$\mathbf{F}_{\alpha}=
\underbrace{\mathbf{F}_{{\rm c},\alpha}}_{{\rm cohesion (fluid/fluid)}} +
\underbrace{\mathbf{F}_{{\rm ads},\alpha}}_{{\rm adhesion
(solid/fluid)}}.$$
For the components we have
$$\mathbf{F}_{{\rm c},\alpha} (\mathbf{x},t) =
-g_{\alpha\bar{\alpha}}\rho_{\alpha}(\mathbf{x},t) \sum_{k} 
\rho_{\bar{\alpha}}(\mathbf{x}+\mathbf{c}_{k}\Delta t,t)
\mathbf{c}_{k} ,$$ $$\mathbf{F}_{{\rm ads},\alpha} (\mathbf{x},t) =
-G_{{\rm ads},\alpha} \rho_{\alpha}(\mathbf{x},t) \sum_{k}
s(\mathbf{x} + \mathbf{c}_{k} \Delta t, t)\mathbf{c}_{k}$$ with $s = 1$
if there is a surface in the direction of motion and $s=0$ if not.
In proportion to these forces, the surface tensions can be calculated in
dependence on the density gradient as arithmetic average of minimum and
maximum density.
$$\gamma_{\alpha\bar{\alpha}} =g_{\alpha\bar{\alpha}}\left[(\rho_{\alpha} -\rho_{\bar{\alpha}} ) / 2 \right]$$
$$\gamma_{\alpha} = G_{{\rm ads},\alpha} = g_{\alpha\bar{\alpha}} \cdot \eta_{\rm wall} $$
From this we obtain the Young-Laplace law
\begin{equation}\label{huangSigmaLG}
\cos \theta = \frac{G_{{\rm ads},\alpha} - G_{{\rm ads},\bar{\alpha}}}{g_{\alpha\bar{\alpha}}\frac{\rho_{\alpha}-\rho_{\bar{\alpha}}}{2}}.
\end{equation}
In our case we use the same coupling for fluid-fluid and fluid-solid interactions.
Therefore, this equation only depends on the density gradients. However,
the dependence on the coupling parameter $g_{\alpha\bar{\alpha}}$ enters
implicitly.
Also for this method we compare the results to the geometrical
measurement of the contact angle. As before, Fig.~\ref{fig:hngCAreldiff}
shows the deviation of the contact angle (in percent) from the values
observed from the reference measurement for an initial droplet size of
$\approx 70^{3}$ $l.u.^3$. 
The deviation is proportional to the absolute value of the coupling and
decreases for low  $g_{\alpha\bar{\alpha}}$ already at a wettability of
$0.2$. This allows to assume a dependence on the interface thickness.

The deviations of Huang's approach compared to the geometrical measurements are up to 15 percent. Since the
validity was only postulated for a limited set of parameters, i.e.
$g_{\alpha\bar{\alpha}}\cdot\rho^{\alpha}=\textrm{const.}$, there might be a range where deviations are lower.
Further, due to the implicit dependence on the coupling, we expect that it should
be possible to achieve a better agreement  of theory and simulation if one
tunes the parameters consistently. However, this is beyond the scope of the
current contribution.

\section{Discussion and conclusion}
We studied the dependence of the contact angle of a droplet on a hydrophobic
surface by means of the Shan-Chen multicomponent LB model and our fluid-surface
interaction model.

First, geometrical measurements of the contact angle were used to measure
parameter dependencies. Parameters taken into consideration here were system
size, coupling parameter $g_{\alpha\bar{\alpha}}$ and wetting parameter
$\eta_{\rm wall}$.
The influence of the system
size on the simulations is caused by finite size effects only and
vanishes when simulating larger systems. Discretization errors for curved
surfaces diminish then, as well as effects of strictly local force
incorporation, leading for instance to finite interfacial thickness.

The pseudo density $\eta_{\rm wall}$ of the wall component was introduced into
the model as parameter of the wetting
behaviour~\cite{bib:jens-kunert-herrmann:2005}. Resonably far from the extremal
cases of complete (de-)wetting ($\theta = 0^{\circ}$ and $\theta =
180^{\circ}$, respectively), a linear dependency of $\theta$ on $\eta_{\rm
wall}$ was observed. This behaviour can be understood following the concept of
Eq. (\ref{huangSigmaLG}). The coupling parameter $g_{\alpha\bar{\alpha}}$ of
the intercomponent interaction is the same for all components (fluid-fluid as
well as fluid-wall) and therefore cancels from the Young-Laplace law, leaving
the contact angle proportional to the ratio of densities only.
Nonetheless, since the coupling parameter $g_{\alpha\bar{\alpha}}$ is
determining the density gradient at the interfacial area, there is still an
indirect influence on the contact angle. Here, two
effects can be differentiated. Given lower coupling, the interfacial area
becomes more diffuse, introducing a higher uncertainty to the determination of
the position of the interface. For high values of $g_{\alpha\bar{\alpha}}$ and
thus strong repulsive forces, the pseudo potential of the wall can cause the
droplet to hover, thereby leaving the definition range of the contact angle.

A method to calculate the expected contact angle as a function of parameters
would be expedient. A first ansatz to deduct the contact angle of a
single phase multicomponent system from a simple model is given by the
determination of the surface tension between each two of the three components
present in the droplet system. The main advantage of this approach lies in the
small system size needed and the possibility to tabulate the obtained values
for future use. Because of the periodic boundaries the precision of the
calculation is relying only on the dimension normal to the
interface~\cite{bib:gonzalez-coveney-04}. The surface tension is then
determined by its mechanical definition, Eq.~(\ref{eq:gammaMechDef}).
Comparison between the contact angles calculated by inserting these surface
tension values into the Young-Laplace law and the ones measured geometrically
in droplet systems yields however large discrepancies. While the range of
definition is met for coupling parameters close to numerical instability, in
general the contact angle values gained from the model system are much higher
than those observered in the droplet system, reaching the complete dewetting
limit comparably faster.

To quantify the effect of the simplifications made in the model system, mainly
by neglecting the presence of the minority component in the interfacial area as
well as the curvature of the interface, the principal of measurement was
utilised directly in droplet systems as well. The range of definition of
$\theta$ was met for the whole coupling parameter range. However, in the
range of linear $\eta_{\rm wall}$-dependence found by geometrical measurement
the contact angle calculated from the Young-Laplace equation was in
general lower, diverging by up to $18\%$. 

A problem still persisting with surface tension measurements in the droplet
system are discretization effects at the curved interface. Addionally, the
interfacial range where there is actual tensorial pressure, is depending on the
coupling chosen, 5 to 11 lattice units wide. This introduces a large
uncertainty to the integration even along the interface orthogonal to lattice
directions. 
Whether a tuning of the contact angle behaviour by introducing
separate coupling for each two components is possible is yet to be determined.

Finally, to evaluate another approach of a priori contact angle determination,
an approximation introduced by Huang et al. for a multiphase multicomponent
Shan Chen model was adapted to our single phase multicomponent approach. Here
results comparable to the surface tension measurements in the droplet system
were gained. While the range of definition is met, the contact angle values are
up to $15\%$ lower than the geometrically measured. However, since the
approximation was postulated for a fixed relation of density and coupling,
eventually a change in the parameter set can decrease this deviation. Because
of the high calculation cost of parameter search this has been omitted.

To conclude, utilising a pseudo wall density as wetting parameter in a single phase
multicomponent Shan Chen LBM it is possible to simulate the complete range of
contact angles as determined by geometrical measurement. A priori determination
of the contact angle based on simulation parameters is possible with an
uncertainty between 10 and 20\% depending on the schemes taken into
consideration as well as the parameter range. 

\section*{Acknowledgments}
This work was supported by the DFG priority program ``nano- and microfluidics''
and the collaborative research centre (SFB) 716. The computations were
performed at the J{\"u}lich Supercomputing Centre and the Scientific
Supercomputing Centre Karlsruhe.


\end{document}